\documentclass[useAMS,usenatbib]{mn2e}
\usepackage{graphicx}
\usepackage{amsmath}
\usepackage{array}
\usepackage{multicol}
\usepackage{multirow}
\usepackage{longtable}
\usepackage[colorlinks=true,
            linkcolor=blue,
            urlcolor=blue,
            citecolor=blue]{hyperref}

\title[Dilution factors of SE-SNe: Cosmology]{Empirically determined dilution factors of stripped-envelope, core-collapse SNe: Paper II - Using GRB-SNe to determine the Hubble Constant}

\author[Cano]{\noindent Zach Cano$^{1,2,3}$\thanks{zewcano@gmail.com}   \\
\noindent $^1$Centre for Astrophysics and Cosmology, Science Institute, University of Iceland, Dunhagi 5, 107 Reykjavik, Iceland.\\
\noindent $^2$Instituto de Astrof\'isica de Andaluc\'ia (IAA-CSIC), Glorieta de la Astronom\'ia s/n, E-18008, Granada, Spain.\\
\noindent $^3$Juan de la Cierva Fellow.\\
}

\begin{document}

\date{Accepted xx. Received xx; in original form xx}

\pagerange{\pageref{firstpage}--\pageref{lastpage}} \pubyear{2013}

\maketitle

\label{firstpage}

\begin{abstract}
The aim of this work is to use gamma-ray burst supernovae (GRB-SNe), which have a luminosity$-$decline relationship akin to that of SNe Ia, as cosmological probes to measure the Hubble constant, $H_0$, in the local Universe. In the context of the Expanding Photosphere Method (EPM), I use empirically derived dilution factors of a sample of nearby SNe Ic, which were derived in Paper I of a two-paper series, as a proxy for the dilution factors of GRB-SNe.  It is seen that the dilution factors as a function of temperature in $VI$ display the least amount of scatter, relative to $BVI$ and $BV$.  A power-law function is fit to the former, and is used to derive model dilution factors which are then used to derive EPM distances to GRB-SNe 1998bw and 2003lw: $36.7\pm9.6$ and $372.2\pm137.1$~Mpc, respectively.  In linear Hubble diagrams in filters $BVR$, I determine the offset of the Hubble ridge line, and armed with the peak absolute magnitudes in these filters for the two aforementioned GRB-SNe, I find a (weighted average) Hubble constant of $\bar{H_{0,\rm w}} = 61.9\pm12.3$~km~s$^{-1}$~Mpc$^{-1}$ for GRB-SNe located at redshifts $z\le0.1$.  The 20\% error is consistent with the value of $H_0$ calculated by Planck and SNe Ia within 1$\sigma$.  I tested the fitting method on five nearby SNe Ic, and found that their EPM distances varied by 18-50\%, with smaller errors found for those SNe which had more numerous usable observations.  For SN 2002ap, its EPM distance was overestimated by 18\%, and if the distance to SN~1998bw was similarly over-estimated by the same amount, the resultant value of the Hubble constant is $H_0 = 72$~km~s$^{-1}$~Mpc$^{-1}$, which perfectly matches that obtained using SNe Ia.  In the conclusions I outline a GRB-SN campaign that can fully exploit GRB-SNe as cosmological probes, and result in additional distances to their host galaxies, which may ultimately result in the usage of GRBs themselves as cosmological probes over almost all of cosmic time.
\end{abstract}

\begin{keywords}
TBC
\end{keywords}

\section{Introduction}

In 2014 it was demonstrated that gamma-ray burst supernovae, GRB-SNe \citep{WoosleyBloom06,CanoReview17} have observed relationships between their absolute peak brightness and the shape of their optical light curves (LCs): a luminosity$-$stretch relation \citep{Cano14} and an analogous luminosity$-$decline relation (LDR; \citealt{CJG14}, CJG14 hereon, and \citealt{LiHjorth14}).  The amount of scatter in the GRB-SNe $BVR$ LDRs was of order $\sigma = 0.2-0.3$~magnitudes.  That GRB-SNe have a LDR akin to that of SNe Ia \citep{Phillips1993} suggests that they too can be used as cosmological probes.  Initial analyses showed that low-redshift ($z\le0.2$) GRB-SNe can be used \emph{in principle} to constrain the value of the Hubble constant ($H_{\rm 0}$) in the local universe (CJG14), and provide estimates of the mass ($\Omega_{\rm M}$) and energy densities ($\Omega_{\rm \Lambda}$) of the universe \citep{Li14}.  In the low-redshift GRB-SNe Hubble diagrams of CJG14, the amount of scatter was $\sigma = 0.3$~mag in the $B$-band, which is precisely the same amount of scatter measured for SNe Ia over the same redshift range \citep{Betoule14}.


Using GRB-SNe as cosmological probes has obvious drawbacks: they numbers are very small (only 46 published GRB-SNe from 1998$-$2016; \citealt{CanoReview17}), compared with the thousands of SNe Ia observed to date.  GRB-SNe have been detected up to a redshift of unity, whereas SNe Ia have been observed up to $z=2$ \citep{Jones13}.  However, with the (hopefully successful) launch of NASA's James Webb Space Telescope (JWST), it will be possible to detect GRB-SNe up to redshifts of $z=3-5$ (CJG14).  Moreover, finding GRB-SNe in the early universe is quite trivial compared with finding SNe Ia at the same distances $-$ one only needs to wait until a high-redshift GRB is detected via its gamma-ray emission, and localized by its X-ray afterglow.

Despite these drawbacks, the future use of GRB-SNe and GRBs as cosmological probes is quite bright.  If relationships can be obtained between GRB-SNe and their high-energy emission, then the large redshift range that GRBs current probe ($0.0086 \le z \le 9.4$; \citealt{Galama98,Cucchiara11}) clearly make them attractive candidate cosmological probes \citep{Schaefer07,Izzo15}.  However, for GRBs to successfully provide constraints on cosmological models without introducing a ``circularity problem'' (where a cosmological model must first be assumed to obtain distances to GRBs from their redshifts), cosmological-model-independent distances to GRBs and/or their host galaxies need to be obtained.  

Overcoming this hurdle is the main focus of this paper: obtaining model-free distances to GRB-SNe and their host galaxies.  There are many ways to achieve this goal, and when considering the advantages and drawbacks of each method, I decided upon a kinematic approach referred to as the Expanding Photosphere Method (EPM).  Here, in Paper II of a two-paper series, I demonstrate how the dilution factors as a function of temperature ($\zeta(T)$) of a sample of nearby SNe Ic that were empirically derived in Paper I, can be used as a proxy for GRB-SNe.  In Section~\ref{sec:methods} I quickly recap on how I assembled the sample of SNe, while in Section~\ref{sec:EPM_background} I present the theoretical derivation of the EPM. In Section~\ref{sec:dilution_factor} I obtain model dilution factors for the GRB-SN sample, and use them to calculate their EPM distance in Section~\ref{sec:GRBSNe_distances}.  Hubble diagrams of GRB-SNe in filters $BVR$ are presented in Section~\ref{sec:Hubble_diagrams}, and ultimately calculate the Hubble constant in Section~\ref{sec:H0}. I discuss the limitations and caveats of the method in Section~\ref{sec:caveats}, and finally present the conclusions in Section~\ref{sec:conclusions}.

\section{Methods}
\label{sec:methods}

\subsection{The Supernova sample}

The criteria for including a given GRB-SN in the sample is described in detail in Paper I.  Briefly for completeness, the main features are:

\begin{itemize}
 \item Observations in two or more filters that bracket a rest-frame $BVRI$ filter(s).
 \item The magnitudes must be host-subtracted (either mathematically or via the image-subtraction technique).
 \item Knowledge of the entire line-of-sight extinction (both local to the SN \& from the Milky Way).
 \item The LC in a given filter must be sampled well enough that an estimate of its peak time, peak magnitude and the $\Delta m_{15}$ parameter\footnote{i.e. the amount the light curve fades in a given filter from peak light to fifteen days later} can be determined.
\end{itemize}

Once SNe with suitable observations were identified in the literature, I undertook the following general steps to obtain decomposed K-corrected LCs of each SN:

\begin{enumerate}
 \item Remove the host contribution.
 \item Correct for foreground extinction.
 \item Convert magnitudes into monochromatic fluxes using zeropoints in \citet{Fukugita95}.
 \item Create observer-frame spectral energy distributions (SEDs) and interpolate to $BVRI$~$(1+z)$ wavelengths and extract the flux.
 \item Correct for rest-frame extinction.
\end{enumerate}

In addition to the above steps, the GRB-SN LCs were further decomposed to isolate the flux coming from the SN itself, using the method presented in Paper I.

\subsection{Estimation of observable parameters}

Using the observations obtained via the method described in the preceding section, the resultant SN LCs were modelled to determine the peak apparent magnitudes in each filter. Three different functions were used to determine these observables: (1) the Bazin function \citep{Bazin11}, (2) high-order polynomials, and (3) linear splines.  All functions were fit to the data using \textsc{python} scripts, including the linear-least-squares (LLS) Levenberg-Marquardt algorithm via \textsc{scipy.optimize.curve$\_$fit}, \textsc{numpy.polyfit} and \textsc{scipy.interpolate.interp1d}.

Note that I adopted the conservative 20\% error estimated by C14 for the peak magnitudes and $\Delta m_{15}$ values of the GRB-SNe.  This conservative error arises from the fact that the isolation of the SN light involves a complicated decomposition technique which contains several sources of error that arise from GRB afterglow modelling and subtraction, the host-galaxy subtraction, the uncertainties in the extinction (both from sight-lines through the Milky Way, and from the SN's host galaxy), and the SED interpolation.  The 20\% error adopted here is much larger than the systematic error associated with the different fitting functions described above.  I have estimated and propagated errors in the analysis for the remaining SNe in the sample in a similar fashion, which are smaller than those of the GRB-SNe due to the fact that there are no uncertainties associated with the AG modelling and removal. Note that this conservative error does not apply to GRB-SNe that did not have a strong afterglow component, including SN~1998bw, and SNe 2006aj and 2010bh.

Finally, the LLS Levenberg-Marquardt algorithm (\textsc{scipy.optimize.curve$\_$fit}) was used to fit the Hubble ridge lines in Section~\ref{sec:Hubble_diagrams} to determine the offset ($\delta$).  All errors quoted here are statistical in nature.

\begin{figure*}
 \centering
 \includegraphics[width=\hsize]{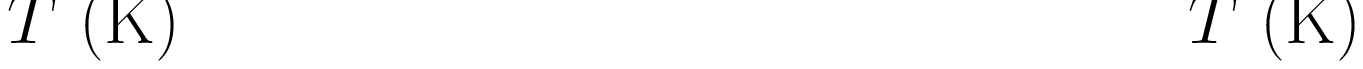} 
 \caption{\textit{Top row}: Empirically derived dilution factors ($\zeta$) in filter combinations ($BVI$, $BV$ and $VI$, left to right) of SNe IIb (red), SNe Ib (blue), SNe Ic (green) and the two relativistic SNe IcBL (black) in the sample as a function of temperature (from Paper I). Each sub-type is respectively presented in the following rows. It is seen that the largest amount of scatter in seen in the $BV$ filter combination, where the dilution factors of the entire SE-SN sample span roughly an order of magnitude at a given temperature.}
 \label{fig:DF_all}
\end{figure*}

\begin{figure*}
 \centering
 \includegraphics[width=\hsize]{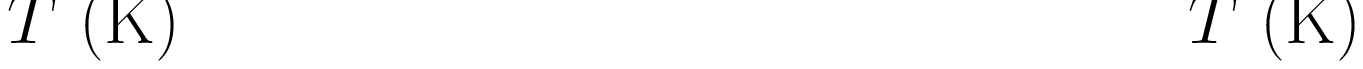} 
 \caption{\textit{Top row}: Empirically derived dilution factors ($\zeta$) in filter combinations ($BVI$, $BV$ and $VI$, left to right) of SNe IIb (red), SNe Ib (blue), SNe Ic (green) and the two relativistic SNe IcBL (black) in the sample as a function of temperature over the time range $0.2 \times t_{\rm bolo}^{\rm max} < t < 1.1 \times t_{\rm bolo}^{\rm max}$. Each sub-type is respectively presented in the following rows. }
 \label{fig:DF_short}
\end{figure*}

\section{The Expanding Photosphere Method}
\label{sec:EPM_background}

The ultimate aim of this work is the derivation of cosmological-model-independent distances to GRBs and GRB-SNe, and hence their unbiased use as cosmological probes.  There are many methods in which to obtain independent distances: monitoring extragalactic Cepheid variable stars, the planetary nebula luminosity function, surface-brightness fluctuations, the diameters of inner ring structures in S(r) galaxies, the spectra-fitting expanding atmosphere method, the SN II standard candle method, the Tully-Fisher method (which unfortunately fails for the nearest GRB-SN 1998bw; \citealt{Arabsalmani15}), the brightnesses of the brightest red and blue supergiants, distances determined from the concept of sosie galaxies, using the distances to SNe Ia that have occurred in the same galaxy, and the use of theoretical criteria for the gravitational stability of gaseous disks.  

In this work, I used a kinematic model to relate the change in radius of the expanding photosphere to an absolute distance.  This method is known as the EPM, which is a variant of the Baade$-$Wesselink method (Baade 1927), and provides a framework in which to compare the angular size of the photosphere with its measured expansion velocity.  The theoretical schema has been discussed extensively in the literature (e.g. \citealt{Kirshner74,Schmidt92,Schmidt94,Eastman96,Vinko04,DH05}, DH05 hereafter, and \citealt{D15}, D15 hereafter), but in order to clarify the concepts, the key features are repeated here.  

For a spatially resolved explosion (i.e. a nova, SN, etc.), its distance ($D$) can be determined from its photospheric radius ($R$) and its apparent angular size ($\theta$) via:

\begin{equation}
  \theta \equiv R_{\rm phot} / D = [v_{\rm phot} (t - t_0) + R_{\rm 0}] / D \approx [v_{\rm phot} (t - t_0)])/D
  \label{equ:theta}
\end{equation}

\noindent where $v_{\rm phot}$ is the photospheric velocity (km s$^{-1}$), $t_{0}$ is the explosion date, which like $t$, is in units of days, and $R_{0}$ is the initial radius of the progenitor.  For all but the very earliest moments, $R >> R_{0}$, and so the latter is neglected hereon\footnote{The radius of a ``typical'' Wolf-Rayet star, the suspected progenitor of GRB-SNe, and possibly SNe Ic-BL too, is of order 5$-$10 $R_{\odot}$ (Crowther 2007), where the higher estimate is $\approx10^{11}$ cm in round numbers.  The early ejecta travels at velocities in excess 40,000 km s$^{-1}$, and covers the distance of a single radius in $\approx$4 min, while at a day will have reached a distance of $>300$ radii.  As the times considered here are $t-t_{0}>2.5$ d, neglecting the progenitor's radius is justified.}. 

For a cosmologically nearby\footnote{Novae/SNe that are sufficiently nearby that the observations do not need a cosmological K-correction $-$ i.e. the correction that arises from observing events occurring at vast cosmological distances, where the observer-frame light arises from bluer rest-frame light that has been redshifted as it traverses through an expanding universe.} ($z<<1$) nova/SN, and one that is a perfect blackbody emitter, the amount of flux $f_{\rm \nu}$ received by an observer at earth is:

\begin{equation}
 f_{XYZ,\rm \nu} = \theta^2 \pi B_{\rm \nu}(T_{XYZ}) 10^{(-A_{\rm \nu} / 2.5)}
 \label{equ:flux}
\end{equation}

\noindent the notation $XYZ$ indicates the flux recorded for different filter combinations (e.g. $BV$, $BVI$, $VI$, etc.), and $A_{\rm \nu}$ is the dust extinction.  $B_{\rm \nu}(T_{XYZ})$ is the Planck function at color temperature $T_{XYZ}$, defined as:

\begin{equation}
 B_{\rm \nu}(T_{XYZ}) =  \Lambda \frac{2h\nu^3}{c^2 }\left[{\rm exp}\left(\frac{h\nu}{k_{\rm B}T_{XYZ}}\right) -1\right]^{-1}
 \label{equ:BB}
\end{equation}


\noindent where $R_{\rm BB}$ is the radius of the blackbody emitter, and $\Lambda$ is a normalization constant to convert between SI and cgs units ($\Lambda = 10^{26}$ to convert between W~m$^{-2}$~Hz$^{-1}$ and Jy).

Equating \ref{equ:theta} and \ref{equ:flux}, we find the distance to be:

\begin{equation}
 D = v_{\rm phot} (t - t_0) \left(\sqrt{\frac{f_{XYZ,\rm \nu}}{\pi B_{\rm \nu}(T_{XYZ}) 10^{(-A_{\rm \nu} / 2.5)}}}~\right)^{-1}
 \label{equ:EPM_distance}
\end{equation}

Thus, for a perfect blackbody emitter, the distance can be determined by: (1) modelling the dereddened broadband spectral energy distributions (SEDs) with the Planck function to determine the radius of the blackbody emitter as a function of time; (2) find the photospheric radius as a function of time by using the blueshifted velocities of one or more line transitions in the optical/NIR spectra as a proxy.  In this work I use the blueshifted velocity of the Si \textsc{ii} $\lambda$6355 transition for the SNe Ic in the sample.  Caveats of using using a single line transition as a proxy for the photospheric velocity are discussed extensively in DH05, as well in Paper I.

\section{The Dilution Factor}
\label{sec:dilution_factor}

However, novae/SNe are not perfect blackbody emitters.  Indeed, the formalization presented in the previous section assumes that the thermalization radius is exactly equal to the photospheric radius (defined as the location where the total inward-integrated radial optical depth reaches a value of $2/3$, e.g. DH05) where the photons escape into space unimpeded.  In reality, especially during the photospheric phase when the SN ejecta is partially or near fully ionized and electron-scattering is a significant contributor to the optical opacity, the radius of thermalized outflow layer is less than the photospheric layer ($R_{\rm BB} < R_{\rm phot}$).  This implies that there is a global source of thermalized photon dilution, which historically has been called the dilution factor ($\zeta$) or the distance-correction factor, and is introduced to equ. \ref{equ:flux} such that:

\begin{equation}
 f_{XYZ,\rm \nu} = \zeta^2 \theta^2 \pi B_{\rm \nu}(T_{XYZ}) 10^{(-A_{\rm \nu} / 2.5)}
 \label{equ:dilute_flux}
\end{equation}

\noindent thus, for a perfect blackbody $\zeta=1$.  In comparison, the dilution factor for SNe II has been determined via radiative-transfer simulations (e.g. Eastman, Schmidt \& Kirshner 1996; DH05), which has values $\zeta\le0.2$ for strongly ionized models.

The physics underlying $\zeta$ is complex\footnote{A theoretical derivation of the dilution factor is beyond the scope of this work, and instead can be found in e.g. sect. 4 of DH05.}, and its precise value depends on the temperature, composition and density of the SN atmosphere, as well as the thermalization radius.  Indeed a key quantity that regulates the amount of dilution present is the spatial separation between $R_{\rm BB}$ and $R_{\rm phot}$, which is the focus of this work here.  Empirically, we can find $\zeta$ for a given SN if we know its explosion time ($t_0$) and its distance from Earth.  Then, as discussed in the previous section, $R_{\rm BB}$ and $R_{\rm phot}$ can be determined from the dereddened observations, and hence the dilution factor is calculated as the ratio:

\begin{equation}
 \zeta = \frac{R_{\rm BB}}{R_{\rm phot}}
\end{equation}

Strictly speaking, $\zeta$ represents the amount of correction needed to transform the measured blackbody flux into the observed flux.  So while the varying photospheric and blackbody radii contribute to this correction term, other factors are also at play \textit{including} the dilution of flux arising from the strongly scattering SN atmosphere.  So what is referred to here as the dilution factor can be more appropriately referred to as a blackbody$-$observed flux ``correction'' factor.  Nevertheless, I adopt the former term throughout this paper.

\begin{figure}
 \centering
 \includegraphics[width=\hsize]{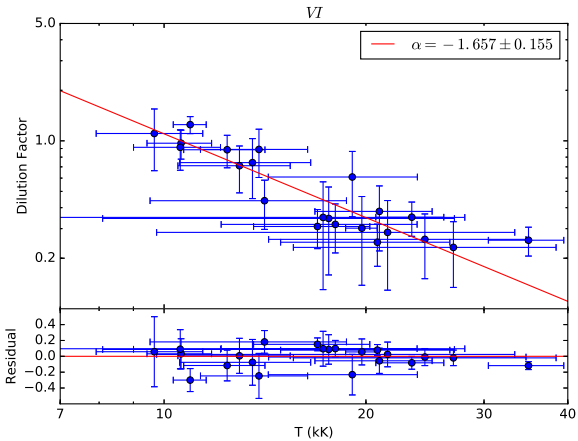}
 \caption{Power-law fit to the dilution factors in $VI$ as a function of temperature. The best-fitting power-law index was found to be $\alpha = 1.657 \pm 0.155$.  Using this best-fitting line, I extracted the model dilution factor at each observed $VI$ temperature for GRB-SNe 1998bw and 2003lw (the only two GRB-SNe where the $VI$ temperatures could be determined) which were ultimately used to determine a distance to its host galaxy.  }
 \label{fig:DF_VI}
\end{figure}

In Paper I, I empirically derived the dilution factors of a sample of SNe Ib, Ic, and IIb from photometric and spectroscopic observations.  Crucially, all of the SNe had cosmological-model-free distances determined for their host galaxies.  Hence the derived dilution factors contain no bias based on any cosmological model, or the values of any cosmological parameter (i.e. the Hubble constant, the mass and/or energy density of the cosmos).  The dilution factors in filter combinations $BV$, $BVI$ and $VI$ found in Paper I are re-shown here in Fig.~\ref{fig:DF_all}.  A quick inspection of the diagrams reveals a lot of scatter in each filter, especially in colours $BV$ and $BVI$.  In an attempt to overcome this scatter, I considered only those dilutions factors which were obtained for the following phases relative to peak bolometric light, as per D15: $0.2 \times t_{\rm bolo}^{\rm max} < t < 1.1 \times t_{\rm bolo}^{\rm max}$.  A plot showing the dilution factors over this shortened time-range is given in Fig.~\ref{fig:DF_short}.  It is seen that a considerable amount of scatter is still present in the combined sample for all three filter combinations, which also applies to the individual SN IIb and Ic samples.  Encouragingly, less scatter is seen in the SN Ic sample in the $VI$ colour, to which I fit a power-law function to determine the normalization constant ($\Sigma$) and the power-law index ($\alpha$), as shown in Fig.~\ref{fig:DF_VI}.  The best-fitting power-law index was found to be $\alpha = 1.657 \pm 0.155$.

\subsection{EPM Distances to two nearby GRB-SNe}
\label{sec:GRBSNe_distances}

Next, I used the best-fitting power-law function to the $VI$ dilution factors found in the previous section to extract model dilution factors as a function of temperature.  Then, for each $VI$ temperature and the model dilution factor, an EPM distance to the host galaxy of SNe 1998bw and 2003lw (the only two GRB-SNe for which the rest-frame $VI$ temperature could be determined) was calculated using equ.~\ref{equ:EPM_distance}.  The results are shown in Table~\ref{table:EPM_distances}, where the quoted errors include not only the propagated observational errors, but also those from the fit.  

In total, I was able to calculate an EPM distance from six measurements of SN~1998bw and three measurements of SN~2003lw.  For each SN I found the weighted average, and its associated error; for SN~1998bw I calculated $\bar{D_{\rm w}} = 36.7\pm9.6$~Mpc, and for SN~2003lw I determined $\bar{D_{\rm w}} = 372.2\pm137.1$~Mpc.  The errors are dominated by the uncertainties in the model dilution factor ($\Delta \zeta$), where a decrease in $\Delta \zeta$ by a factor of 10 would lead to a reduced error in the distance of 25\%.

\begin{table}
\small
\centering
\setlength{\tabcolsep}{5pt}
\caption{Model dilution factors ($VI$) and the calculated EPM distance to GRB-SNe 1998bw and 2003lw}
\label{table:EPM_distances}
\begin{tabular}{|cccc|}
\hline
Name	&		$T_{VI}$ (K)				&		$\zeta_{\rm model}$				&		Distance				\\
\hline																			
1998bw	&	$	17938	\pm	4011	$	&	$	0.418	\pm	0.186	$	&	$	42.9	\pm	23.5	$	\\
1998bw	&	$	21353	\pm	5619	$	&	$	0.313	\pm	0.174	$	&	$	36.0	\pm	24.1	$	\\
1998bw	&	$	21731	\pm	5822	$	&	$	0.304	\pm	0.170	$	&	$	35.2	\pm	23.8	$	\\
1998bw	&	$	21731	\pm	5884	$	&	$	0.304	\pm	0.170	$	&	$	36.6	\pm	24.9	$	\\
1998bw	&	$	20968	\pm	5680	$	&	$	0.330	\pm	0.180	$	&	$	32.4	\pm	22.1	$	\\
1998bw	&	$	18633	\pm	4824	$	&	$	0.393	\pm	0.182	$	&	$	37.4	\pm	23.2	$	\\
\hline
1998bw$^*$	&	$	-			$	&	$	-			$	&	$	36.7	\pm	9.6	$	\\
\hline
2003lw	&	$	23996	\pm	1200	$	&	$	0.258	\pm	0.178	$	&	$	319.2	\pm	222.8	$	\\
2003lw	&	$	18602	\pm	930	$	&	$	0.394	\pm	0.186	$	&	$	444.4	\pm	215.6	$	\\
2003lw	&	$	28615	\pm	1431	$	&	$	0.193	\pm	0.170	$	&	$	330.9	\pm	293.8	$	\\
\hline
2003lw$^*$	&	$	-			$	&	$	-			$	&	$	372.2	\pm	137.1	$	\\
\hline
\end{tabular}
\begin{flushleft}
$^*$The weighted average linear distance and its associated error.\\
\end{flushleft}
\end{table}

\begin{figure*}
 \centering
 \includegraphics[width=\hsize]{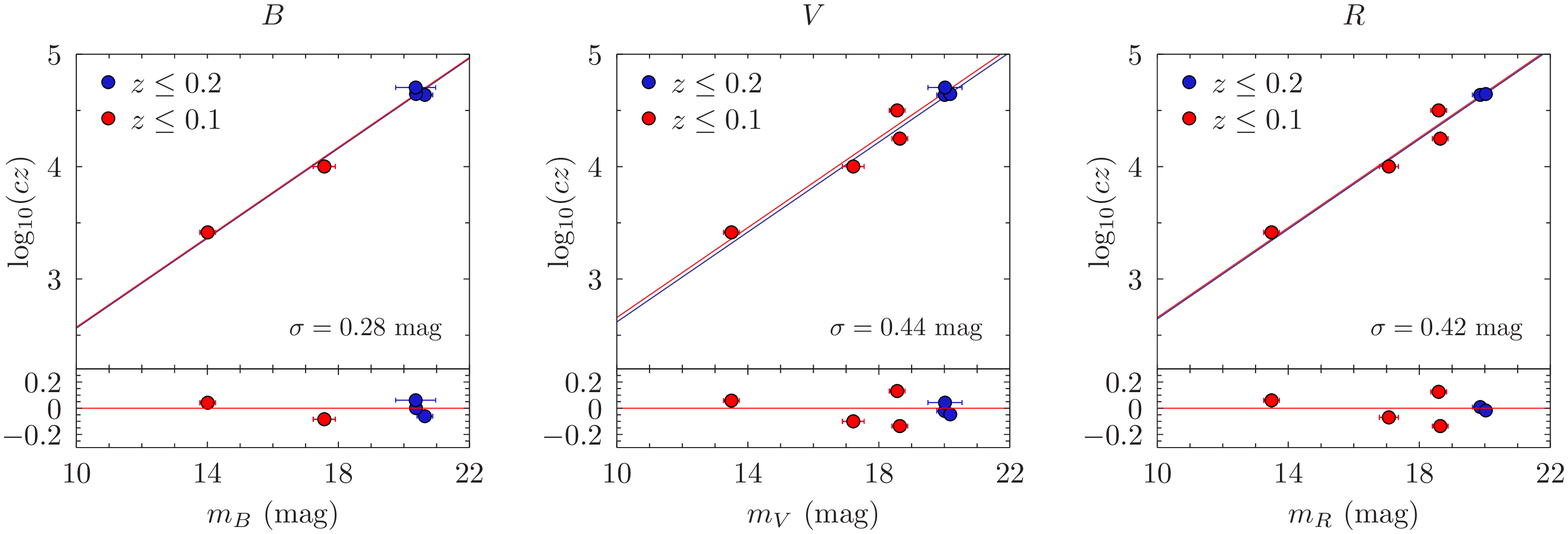}
 \caption{Determining the offset to the linear Hubble ridge line (equ. \ref{equ:hubbleline}) using low-redshift GRB-SNe. Two redshift ranges are considered here: (1) $z\le0.1$ (red), and (2) $z\le0.2$ (blue).  The best-fitting values for the offset are shown in Table~\ref{table:ridge_line}.  The rms ($\sigma$) of the observations to the fitted ridge line for the $z\le0.1$ sample in each filter are shown, while the residuals relative to the $z\le0.1$ ridge line are shown in the bottom panel of each subplot for each filter.  It is seen that the lowest amount of scatter is seen in the bluer $B$ filter ($\sigma\approx0.3$~mag), while the scatter in the redder filters is approximately the same $\sigma\approx0.4$~mag).}
 \label{fig:Hubble_diagram}
\end{figure*}

\section{GRB-SN Cosmology}
\label{sec:cosmo}

\subsection{Linear Hubble diagram of GRB-SNe}
\label{sec:Hubble_diagrams}

We now have almost all the necessary ingredients to use GRB-SNe to measure the Hubble constant, including the peak absolute magnitudes determined from model-free distances.  However, to calculate $H_0$, we first need to construct a Hubble diagram (in the linear, low-redshift regime) and determine its offset ($\delta$).  Equation \ref{equ:hubbleline} gives the Hubble ridge line (e.g. \citealt{Tammann02,ST93}), which is valid for nearby SNe whose redshift are $z\ll1$.  It is important to omit events a larger redshifts as they are no longer in the local Hubble flow and their inclusion in the fit can result in miscalculating the local value of the Hubble constant. For example, SNe Ia studies (e.g. \citealt{Riess04,Riess07}) impose an upper limit of $z\le0.1$ when fitting to obtain $H_{0}$.  Given the relative paucity of SNe in the sample, I have investigated Hubble diagrams for two samples of GRB-SNe: those with $z\le0.1$ and $z\le0.2$.

\begin{equation}
 \mathrm{log}(cz) = 0.2 \times m_{\rm app} + \delta
 \label{equ:hubbleline}
\end{equation}

From the sample of GRB-SNe I found their peak apparent magnitudes in filters $B$, $V$ and $R$ (Table~\ref{table:vitals}) and fit the Hubble ridge line for the two redshift ranges to determine $\delta$ in each filter, as shown in Fig.~\ref{fig:Hubble_diagram}. It is important to recall the small datasets investigated here: in $B$ there are only two at $z\le0.1$ and five at $z\le0.2$, in $V$ there are four at $z\le0.1$ and seven at $z\le0.2$, and finally in $R$ there are four at $z\le0.1$ and six at $z\le0.2$.  The effects of small sample sizes is discussed in Section~\ref{sec:caveats}.  The best-fitting values for each filter and each of the two redshift ranges are presented in Table~\ref{table:ridge_line}.  For comparison, I also give the offset and its error for the sample of SNe Ia from \citet{Betoule14} which, like the GRB-SNe, are uncorrected for an LDR.

It is seen in for the two redshift ranges, the offset values agree within the errorbars for all filters.  However, the errorbars are rather large, which are the direct result of fitting the small samples.  In contrast, the value of the offset from the SN Ia sample differs by 0.012 between the two redshift ranges, where the smaller redshift sample resulted in the larger value.  As the errorbars on $\delta$ from the SN Ia samples are very small (0.001), the difference of 0.012 is 6$\sigma$.  Clearly the choice of redshift range is important, and future studies should consider investigating samples for $z\le0.1$.  Nevertheless, for the sake of completeness, and to demonstrate the difference in the derived value of $H_0$ for both redshift ranges, in this work we will continue to consider both ranges of GRB-SNe.

\begin{table}
\small
\centering
\setlength{\tabcolsep}{4pt}
\caption{Hubble ridge line offset ($\delta$)}
\label{table:ridge_line}
\begin{tabular}{|cccccc|}
\hline
type	&	Filter	&	redshift range	&	$N$	&		$\delta$				&	$\sigma$	\\
\hline															
Ia	&	$B$	&	$z\le0.1$	&	152	&	$	0.643	\pm	0.001	$	&	0.31	\\
Ia	&	$B$	&	$z\le0.2$	&	318	&	$	0.631	\pm	0.001	$	&	0.30	\\
GRB-SNe	&	$B$	&	$z\le0.1$	&	2	&	$	0.571	\pm	0.196	$	&	0.28	\\
GRB-SNe	&	$B$	&	$z\le0.2$	&	5	&	$	0.562	\pm	0.101	$	&	0.28	\\
GRB-SNe	&	$V$	&	$z\le0.1$	&	4	&	$	0.657	\pm	0.128	$	&	0.44	\\
GRB-SNe	&	$V$	&	$z\le0.2$	&	7	&	$	0.617	\pm	0.046	$	&	0.46	\\
GRB-SNe	&	$R$	&	$z\le0.1$	&	4	&	$	0.657	\pm	0.125	$	&	0.42	\\
GRB-SNe	&	$R$	&	$z\le0.2$	&	6	&	$	0.644	\pm	0.053	$	&	0.42	\\
\hline
\end{tabular}
\end{table}

\subsection{$H_{0}$ from low-redshift GRB-SNe}
\label{sec:H0}

Now we have everything we need to calculate the Hubble constant using GRB-SNe.  Next, armed with the absolute $BVR$-band magnitudes of SNe~1998bw and 2003lw, which were calculated using the EPM distances computed in Section \ref{sec:GRBSNe_distances}, and the fitted values of $\delta$ in the $BVR$ Hubble diagrams, we can estimate the Hubble constant via equation \ref{equ:hubbleconstant}:

\begin{equation}
 \mathrm{log}(H_{0}) = 0.2 \times M_{\rm abs} + \delta + 5
 \label{equ:hubbleconstant}
\end{equation}

For the $z\le0.1$ dataset I find the weighted average of the Hubble constant to be $\bar{H_{0,\rm w}} = 61.9\pm12.3$~km~s$^{-1}$~Mpc$^{-1}$, while for the larger $z\le0.2$ dataset I find $\bar{H_{0,\rm w}} = 59.8\pm8.8$~km~s$^{-1}$~Mpc$^{-1}$.  The values clearly agree with each other, which is unsurprising as they have percentage errors of 20\% and 15\%, respectively. These values are within 1$\sigma$ to that determined by Planck ($H_{0} = 67.9\pm0.9$~km~s$^{-1}$~Mpc$^{-1}$; \citealt{Planck16}) and using SNe Ia ($H_{0} = 73.2\pm1.7$~km~s$^{-1}$~Mpc$^{-1}$; \citealt{Riess16}). The 1$\sigma$ agreement clearly arises from the fact that the errors derived from the determination of $H_0$ using GRB-SNe are quite large, which can be reduced in the future with the inclusion of more events.  The smaller error derived from the $z\le0.2$ sample, which contains more GRB-SNe, clearly reflects this.

\begin{table}
\small
\centering
\setlength{\tabcolsep}{4pt}
\caption{$H_0$ calculated using GRB-SNe using the value of $\delta$ from the $z\le0.1$ and $z\le0.2$ samples.}
\label{table:H0}
\begin{tabular}{|cccc|}
\hline
Name	&	Filter	&	$z$ range	&		$H_0$ (km s$^{-1}$ Mpc$^{-1}$)				\\
\hline										
1998bw	&	$B$	&	$\le0.1$	&	$	64.3	\pm	31.9	$	\\
1998bw	&	$V$	&	$\le0.1$	&	$	62.0	\pm	25.4	$	\\
1998bw	&	$R$	&	$\le0.1$	&	$	59.9	\pm	24.3	$	\\
2003lw	&	$V$	&	$\le0.1$	&	$	62.8	\pm	29.3	$	\\
2003lw	&	$R$	&	$\le0.1$	&	$	61.8	\pm	28.6	$	\\
\hline											
Weighted Average	&	-	&	$\le0.1$	&	$	61.9	\pm	12.3	$	\\
\hline											
1998bw	&	$B$	&	$\le0.2$	&	$	63.0	\pm	23.5	$	\\
1998bw	&	$V$	&	$\le0.2$	&	$	56.5	\pm	16.3	$	\\
1998bw	&	$R$	&	$\le0.2$	&	$	61.7	\pm	18.5	$	\\
2003lw	&	$V$	&	$\le0.2$	&	$	57.3	\pm	20.3	$	\\
2003lw	&	$R$	&	$\le0.2$	&	$	63.7	\pm	23.3	$	\\
\hline											
Weighted Average	&	-	&	$\le0.2$	&	$	59.8	\pm	8.8	$	\\
\hline
\end{tabular}
\end{table}

\section{Caveats}
\label{sec:caveats}

\subsection{The effect of the chosen redshift upper limit}

As discussed in Section~\ref{sec:cosmo}, the Hubble ridge line (equ. \ref{equ:hubbleline}) is only valid for small redshifts ($z\ll1$).  It is important to use only the nearest SNe so that determining $H_0$ is not influenced by the energy and mass densities of the universe.  In this work, I reached a compromise of $z\le0.1,$ 0.2 based on the modest sample sizes being considered here.  However we must still ask whether this looser redshift limit will affect the final calculated value of $H_{0}$.

Using the SN Ia sample in \citet{Betoule14}, I fit the ridge line to two samples: (1) $z\le0.2$ and (2) $z\le0.1$.  In case (1) I find $\delta = 0.631$, while in case (2) I find $\delta = 0.643$.  The difference between the two samples amounts to $\Delta \delta = 0.012$.  The effect of the slightly different value of the offset on the value of $H_{0}$ determined for a given SN will have different effects on different SNe.  For example, in equ. \ref{equ:hubbleconstant} we can see that $H_{0}$ is directly proportional to $10^{0.2\times M_{\rm abs}}$, therefore fainter SNe will be more affected by a different value of $\delta$ than brighter SNe. In general, the effect of the looser redshift limit will amount to a difference of 1--4~km~s$^{-1}$~Mpc$^{-1}$ in the final value of $H_{0}$, which is smaller than the errors derived here that are based on the standard deviations calculated for each dataset (which I consider to be 1$\sigma$ errors on the calculated value of $H_{0}$).

\subsection{Effect of small-sample sizes}

In the $B$-band Hubble diagram (left panel in Fig.~\ref{fig:Hubble_diagram}), it was seen that the amount of rms scatter about the fitted Hubble ridge line was commensurate with that determined for the LDR-uncorrected SN Ia sample (Table~\ref{table:ridge_line}), i.e. about 0.3~mag.  The similarities in their scatter hints that GRB-SNe are as accurate as SNe Ia for use as cosmological probes.  However, we are forced to question if the perceived low values of the scatter of the GRB-SNe arises from their small sample sizes $-$ i.e. if these sample sizes were the same as the SNe Ia, would they still exhibit a low amount of scatter, or would the rms values increase?

To try and address this uncertainty, I performed an analysis whereby I randomly selected (using Monte-Carlo sampling) events from the SN Ia sample over a range of sample sizes ($N=4-15$) and fit the ridge line to the smaller dataset, and calculated the rms associated with the fit.  In each fit, the rms value of the small, randomly selected SN Ia sample were in the range $\sigma=0.30-0.35$ mag, which is fully consistent with the rms scatter calculated for the full SN Ia sample.

If the range of scatter in the smaller SN Ia samples were smaller, then we could conclude that the small scatter seen for the combined GRB-SN \& SN IcBL sample was merely a consequence of the small sample sizes used here, and that the actual scatter of the entire relativistic SN population is intrinsically higher.  Instead, the fact that we found similar values for the rms scatter of the smaller SN Ia samples implies that the scatter derived for the small GRB-SN \& SN IcBL sample modelled here can be thought to be representative of its entire population.  This result has encouraging potential for their future use to constrain cosmological models over a larger redshift range to the same degree of precision as SNe Ia.

\subsection{How well do the model dilution factors reproduce the distances to the sample SNe Ic?}
\label{sec:2000ap_test}

One way to test the validity of the approach of using the empirically derived dilution factors of the local SNe Ic as a proxy for the GRB-SNe 1998bw and 2003lw is to use the best-fitting power-law function to determine an EPM distance to a few SNe Ic whose distances are well determined.  I calculated EPM distances for five SNe Ic in the sample: SNe~1994I, 2002ap, 2007gr, 2009bb and 2012ap.  For each of these SNe Ic I followed the sample procedure as that for the GRB-SNe, and hence calculated an EPM distance for each based on the model dilution factor for a given $VI$ colour temperature.  The following list gives a summary of the results: weighted averages are given, and the reference distances are taken from table B1 in Paper I:

\begin{itemize}
 \item \textbf{SN 1994I}: For $N=3$ measurements, I find an EPM distance of $D_{\rm EPM} = 4.5\pm0.9$~Mpc. Compared with a distance of $7.6\pm0.1$~Mpc, the EPM distance is 41\% smaller, i.e. about 3$\sigma$ smaller.
 \item \textbf{SN 2002ap}: For $N=9$ measurements, I find an EPM distance of $D_{\rm EPM} = 10.8\pm1.5$~Mpc. Compared with a distance of $9.2\pm0.6$~Mpc, the EPM distance is 18\% larger, both both agree to within 1$\sigma$.
 \item \textbf{SN 2007gr}: For $N=8$ measurements, I find an EPM distance of $D_{\rm EPM} = 8.3\pm0.7$~Mpc. Compared with a distance of $10.1\pm0.7$~Mpc, the EPM distance is 18\% smaller, or about 2$\sigma$ smaller.
 \item \textbf{SN 2009bb}: For $N=1$ measurement, I find an EPM distance of $D_{\rm EPM} = 60.6\pm16.1$~Mpc. Compared with a distance of $40.7\pm4.4$~Mpc, the EPM distance is 49\% larger, with is about 1$\sigma$ larger, given the large errorbars.
 \item \textbf{SN 2012ap}: For $N=1$ measurement, I find an EPM distance of $D_{\rm EPM} = 18.0\pm8.9$~Mpc. Compared with a distance of $40.4\pm7.4$~Mpc, the EPM distance is 55\% smaller, i.e. about 2$\sigma$ smaller.
\end{itemize}

For three of the five SNe Ic, the EPM method under-estimated the actual distance by between 18-55\%, whereas for relativistic SNe IcBL 2009bb and SN 2002ap, their EPM distances were about 49\% and 18\% larger, respectively.  Indeed for both of the relativistic SNe IcBL, the EPM distance differed to the distance given in table B1 in Paper I by roughly 50\% (one larger, one smaller), which highlights the limited effectiveness of this method.  

Interestingly, for the two SNe with several $VI$ colour temperatures (SNe 2002ap and 2007gr), the EPM distance agrees better with the actual distance to within $\sim$18\%.  This indicates that several $VI$ colour temperature measurements are needed to obtain a more reliable EPM distance.  Investigating further, let us take a closer look at SN~2002ap, whose distance has been well determined, with a weighted average of $D_{\rm w} = 9.2\pm0.6$~Mpc (see table B1 in Paper I). In comparison, I found a weighted average EPM distance of $D_{\rm w} = 10.8\pm1.5$~Mpc.  These are clearly consistent within its errorbars with the weighted average distance to its host, M74, found via other methods (e.g. the planetary nebula luminosity function, tip of the red giant branch method, the EPM and standard candle method for SN II 2013ej, and the Tully Fisher method).  However, let us ignore the errorbars for a moment: its weighted average EPM distance is roughly 18\% larger than that from collected distance calculated via the other methods.  If the distance to the host galaxy of SN~1998bw was similarly over-estimated by 18\%, its reduced distance would be 31.1~Mpc.  In turn this implies a distance modulus of 32.46~mag, opposed to 32.82~mag from before, which means a difference of 0.36~mag.  This smaller distance implies a lower absolute magnitude and hence a larger Hubble constant: for example, in the $V$-band its peak magnitude would be only $M_V = -18.96$, which with the value of $\delta_V = 0.654$, gives a Hubble constant of $H_0 = 72$~km~s$^{-1}$~Mpc$^{-1}$, i.e. perfectly matching that obtained with SNe Ia.

\section{Conclusions \& Future Prospects}
\label{sec:conclusions}

In this work, and in the context of the EPM, I used the dilution factors of a sample of nearby SNe Ic (whose distances were known independent of any cosmological model) derived from observations as a proxy for the dilution factor of GRB-SNe.  In the plots of dilution factor as a function of temperature, it was seen that the smallest amount of scatter was present in the $VI$ colour.  I fit a power-law function to the dataset, finding a best-fitting power-law index of $\alpha = 1.657\pm0.155$.  Then, for GRB-SNe 1998bw and 2003lw, I extracted the model dilution factor for a given $VI$ temperature determined from observations obtained during phases $0.2 \times t_{\rm bolo}^{\rm max} < t < 1.1 \times t_{\rm bolo}^{\rm max}$, and used it to calculate its EPM distance. Using this method I found (weighted average) distances of $36.7\pm9.6$ and $372.2\pm137.1$~Mpc, respectively.  

In linear Hubble diagrams, I determined the offset ($\delta$) of the Hubble ridge line (equ.~\ref{equ:hubbleline}) in filters $BVR$. It was seen that the amount of scatter in the $B$-band was approximately the same for the GRB-SN sample and the sample of SNe Ia investigated here, i.e. about 0.3~mag.  I then used the fitted offsets with the absolute magnitudes of the two aforementioned GRB-SNe, which allowed us to calculate the Hubble constant for two redshift ranges: $z\le0.1$,0.2, the larger range being considered due to the paucity of GRB-SNe in the sample.  I found weighted averages of $\bar{H_{0,\rm w}} = 61.9\pm12.3$~km~s$^{-1}$~Mpc$^{-1}$, and $59.8\pm8.8$~km~s$^{-1}$~Mpc$^{-1}$, respectively, which agree with those found by Planck and SNe Ia within 1$\sigma$ (which is not unsurprising given the large errors on $H_0$ found using GRB-SNe: 20\% and 15\% for the two samples, respectively).

Note that the peak magnitudes of the GRB-SNe used in the linear Hubble diagrams are uncorrected for an LDR.  In order for such a correction to be applied, one needs an independent dataset of GRB-SNe whose model-independent absolute magnitudes have been determined.  Clearly none exist at this stage.  Encouragingly, that GRB-SNe and SNe Ia that are both uncorrected for a LDR have a similar amount of scatter in their linear Hubble diagrams ($\approx0.3$~mag) implies the promising use of the former in future studies where it is possible to correct for an LDR prior to their usage.

I also used the method to determine EPM distances to several of the SNe Ic in the sample. For SNe~2002ap and 2007gr, for which they had the most $VI$ colour temperature measurements, I found that EPM distances that agreed with their actual distances to within 18\%.  For SNe~1994I and 2012ap, the discrepancy was larger, roughly 40$-$50\%, while the EPM distance found for SN~2009bb was 50\% larger. For the two relativistic SNe IcBL in the sample, the EPM method is clearly very limited in its ability to determine an accurate distance to each.  For nearby SN Ic 2002ap I found an EPM distance of $10.8\pm1.5$~Mpc, which is approximately 18\% more distant than that calculated for its host galaxy using a variety of different methods, but consistent within 1$\sigma$.  If my method systematically imparted an $\sim18$\% increase in the derived distances to GRB-SNe 1998bw and 2003lw, an 18\% decrease in their distances translates into smaller absolute magnitudes and a larger value of the Hubble constant. Indeed, as shown in Section~\ref{sec:2000ap_test}, an 18\% decrease in the distance to SN~1998bw implies a Hubble constant of 72~km~s$^{-1}$~Mpc$^{-1}$, which would then be in perfect agreement with that obtained using SNe Ia.

Moving forward, to better utilize the EPM method described here, which is based on $VI$ colour temperatures measured for phases $0.2 \times t_{\rm bolo}^{\rm max} < t < 1.1 \times t_{\rm bolo}^{\rm max}$ relative to peak light, more local SNe Ic are needed whose host galaxy distances have been determined independent of any cosmological model.  More datapoints will help improve the statistical significance of the fitted power-law equation, reducing the errorbars in the fitted parameters and ultimately decreasing the uncertainties in the derived EPM distances.  Additional SNe Ic will also help identify if all SNe Ic follow the same $\zeta(T)$ relationship, or whether more than one relationship is present.

In conclusion, I have presented a case study whereby it is possible to use GRB-SNe to constrain the Hubble constant to an accuracy of $15-20\%$. This accuracy can be reduced simply by the inclusion of more events: more local SNe Ic to better constrain $\zeta(T)$ and more GRB-SNe to better populate the Hubble diagrams.  By no means do I suggest that GRB-SNe are competitive with SNe Ia as cosmological candles $-$ they are simply too rare and many more steps need to be taken to isolate the SN light from a GRB event relative to SNe Ia.  However, I have demonstrated that they are complementary probes, and their continued use may ultimately result in the use of GRBs themselves as cosmological probes: if the distances to three or more GRB-SNe can be determined, whose GRB prompt emission follows the Amati relation \citep{Amati02}, then it is possible to determine the ``intrinsic'' Amati relation, and hence determine the distance to all GRBs that follow the relation.  Moreover, it is possible to extend the GRB-SN Hubble diagram into regimes dominated by the mass and energy components of the cosmos. Indeed, finding GRBs in the distant universe is rather trivial, one needs only to wait for their prompt emission to be detected, and then localized to arc-second precision via its X-ray afterglow/  As discussed in CJG14, JWST will be able to detect and monitor GRB-SNe up to $z=3-5$, which armed with its NIR camera, means it is possible to obtain observations of the necessary rest-frame filters to utilize the method presented here.  Along these lines, in the following subsection I have outlined a possible GRB-SN observational campaign that has been designed to utilize its observational properties as cosmological probes.

\subsection{A suggested GRB-SN observational campaign}

GRB-SNe have been monitored and analysed for almost two decades, and the multitude of published works have collectively gathered a rich dataset that have revealed clues into many facets of their observational and physical nature.  Moving forward, if the intention is to continue to use GRB-SNe as cosmological probes, the devised observational strategy needs to consider what observations are absolutely vital.  These include:

\begin{itemize}
 \item Particular attention is needed to obtain the necessary observations during the time-range $0.2 \times t_{\rm bolo}^{\rm max} < t < 1.1 \times t_{\rm bolo}^{\rm max}$, for which the EPM is most valid.
 \item Time series spectra, where rest-frame Si~\textsc{ii}~$\lambda$6355 or Fe~\textsc{ii}~$\lambda$5169 can be detected and its blueshifted velocity measured.  
 \item Observer-frame filters that bracket \emph{rest-frame} $BVI$. In particular, for an EPM distance to be determined with some certainty, rest-frame $VI$ are needed.  
 \item The obtained photometric and spectroscopic data needs to be decomposed in order to isolate the SN's contribution to the total observed flux.  This involves removing both the afterglow and host contributions.  Hence good cadence and large time-spans are needed.
 \item The entire line-of-sight extinction needs to be determined, both foreground and that arising local to the event.
 \item Finally, accurate spectroscopic redshifts are necessary.
\end{itemize}

To determine an EPM distance to a GRB-SN, one needs rest-frame $VI$ observations and time-series spectra.  With increase distance means that rest-frame $I$ is pushed further and further into the NIR. For example, at $z=0.5$, rest-frame $I$ ($\lambda_{\rm eff} = 8060$~\AA) is redshifted into observer frame $J$-band.  AT $z=1$, rest-frame $I$ is observer-frame $H$-band.  As such, it certainly is possible to obtain the necessary observations, though one must also recall that increased distance results in increasingly fainter SNe, making the detection of faint objects in NIR filters quite challenging, thus necessitating large exposure times.  Moreover, acquiring time-series spectra of $z=1$ GRB-SNe is exceedingly challenging.  Instead, particular focus should be paid to $z\le0.5$ events, of which all the necessary observations can be realistically obtained.

Finally, it should be stressed that the same observational data are needed for nearby SNe Ic in order to better determine the $\zeta(T)$ relationship.  The suggested observational campaign for GRB-SNe is perfectly suited for SNe Ic: the same observations are needed and encouraged, especially around peak light.

\section{Acknowledgments}

I am very grateful for fruitful discussions with Joszef Vink\'o, Palli Jakobsson, Peter Hoeflich, Steve Schulze and Keiichi Maeda regarding the content of this analysis.

ZC acknowledges support from the Juan de la Cierva Incorporaci\'on fellowship IJCI-2014-21669 and from the Spanish research project AYA 2014-58381-P.

\bibliographystyle{mn2e}

\begin{thebibliography}{99}
\bibitem[Amati et al.(2002)]{Amati02} Amati, L., Frontera, F., Tavani, M., et al.\ 2002, A\&A, 390, 81 
\bibitem[Arabsalmani et al.(2015)]{Arabsalmani15} Arabsalmani, M., Roychowdhury, S., Zwaan, M.~A., Kanekar, N., \& Micha{\l}owski, M.~J.\ 2015, MNRAS, 454, L51 
\bibitem[Bazin et al.(2011)]{Bazin11} Bazin, G., Ruhlmann-Kleider, V., Palanque-Delabrouille, N., et al.\ 2011, A\&A, 534, A43 
\bibitem[Betoule et al.(2014)]{Betoule14} Betoule, M., Kessler, R., Guy, J., et al.\ 2014, A\&A, 568, A22 
\bibitem[Cano(2014)]{Cano14} Cano, Z.\ 2014 , ApJ, 794, 121
\bibitem[Cano et al.(2014)]{CJG14} Cano, Z., Jakobsson, P., \& Pall Geirsson, O.\ 2014, (CJG14) arXiv:1409.3570 
\bibitem[Cano et al.(2017)]{CanoReview17} Cano, Z., Wang, S.-Q., Dai, Z.-G., \& Wu, X.-F.\ 2017, Advances in Astronomy, 2017, 8929054 
\bibitem[Cano et al.(2017)]{Cano17} Cano, Z., Izzo, L., de Ugarte Postigo, A., et al.\ 2017, arXiv:1704.05401 
\bibitem[Chakraborti et al.(2015)]{Chakraborti15} Chakraborti, S., Soderberg, A., Chomiuk, L., et al.\ 2015, ApJ, 805, 187 
\bibitem[Crowther(2007)]{Crowther07} Crowther, P.~A.\ 2007, ARA\&A, 45, 177 
\bibitem[Cucchiara et al.(2011)]{Cucchiara11} Cucchiara, A., Levan, A.~J., Fox, D.~B., et al.\ 2011, ApJ, 736, 7 

\bibitem[Dessart \& Hillier(2005)]{DH05} Dessart, L., \& Hillier, D.~J.\ 2005, A\&A, 439, 671 
\bibitem[Dessart et al.(2015)]{D15} Dessart, L., Hillier, D.~J., Woosley, S., et al.\ 2015, MNRAS, 453, 2189 
\bibitem[Drout et al.(2011)]{Drout11} Drout, M.~R., Soderberg, A.~M., Gal-Yam, A., et al.\ 2011, ApJ, 741, 97 
\bibitem[Eastman et al.(1996)]{Eastman96} Eastman, R.~G., Schmidt, B.~P., \& Kirshner, R.\ 1996, ApJ, 466, 911 
\bibitem[Fukugita et al.(1995)]{Fukugita95} Fukugita, M., Shimasaku, K., \& Ichikawa, T.\ 1995, PASP, 107, 945
\bibitem[Galama et al.(1998)]{Galama98} Galama, T. J., et al.\ 1998a, Nature, 395, 670
\bibitem[Izzo et al.(2015)]{Izzo15} Izzo, L., Muccino, M., Zaninoni, E., Amati, L., \& Della Valle, M.\ 2015, A\&A, 582, A115 
\bibitem[Jones et al.(2013)]{Jones13} Jones, D.~O., Rodney, S.~A., Riess, A.~G., et al.\ 2013, ApJ, 768, 166 
\bibitem[Kirshner  \& Kwan(1974)]{Kirshner74} Kirshner, R.~P., \& Kwan, J.\ 1974, ApJ, 193, 27 
\bibitem[Li \& Hjorth(2014)]{LiHjorth14} Li, X., \& Hjorth, J.\ 2014, arXiv:1407.3506 
\bibitem[Li et al.(2014)]{Li14} Li, X., Hjorth, J., \& Wojtak, R.\ 2014, ApJL, 796, L4 
\bibitem[Margutti et al.(2013)]{Margutti13} Margutti, R., Zaninoni, E., Bernardini, M.~G., et al.\ 2013, MNRAS, 428, 729 
\bibitem[Phillips(1993)]{Phillips1993} Phillips, M.~M.\ 1993, ApJL, 413, L105
\bibitem[Pignata et al.(2011)]{Pignata11} Pignata, G., et al.\ 2011, ApJ, 728, 14
\bibitem[Planck Collaboration et al.(2016)]{Planck16} Planck Collaboration, Ade, P.~A.~R., Aghanim, N., et al.\ 2016, A\&A, 594, A13 
\bibitem[Riess et al.(2004)]{Riess04} Riess, A.~G., Strolger, L.-G., Tonry, J., et al.\ 2004, ApJ, 607, 665 
\bibitem[Riess et al.(2007)]{Riess07} Riess, A.~G., Strolger, L.-G., Casertano, S., et al.\ 2007, ApJ, 659, 98 
\bibitem[Riess et al.(2016)]{Riess16} Riess, A.~G., Macri, L.~M., Hoffmann, S.~L., et al.\ 2016, ApJ, 826, 56 
\bibitem[Sandage \& Tammann(1993)]{ST93} Sandage, A., \& Tammann, G.~A.\ 1993, ApJ, 415, 1
\bibitem[Schaefer(2007)]{Schaefer07} Schaefer, B.~E.\ 2007, ApJ, 660, 16 
\bibitem[Schmidt et al.(1992)]{Schmidt92} Schmidt, B.~P.,  Kirshner, R.~P., \& Eastman, R.~G.\ 1992, ApJ, 395, 366 
\bibitem[Schmidt et al.(1994)]{Schmidt94} Schmidt, B.~P.,  Kirshner, R.~P., Eastman, R.~G., et al.\ 1994, ApJ, 432, 42 
\bibitem[Soderberg et al.(2010)]{Soderberg10} Soderberg, A.~M., et al.\ 2010, Nat, 463, 513 
\bibitem[Tammann et al.(2002)]{Tammann02} Tammann, G.~A., Reindl, B., Thim, F., Saha, A., \& Sandage, A.\ 2002, A New Era in Cosmology, 283, 258 
\bibitem[Tully(1988)]{Tully88} Tully, R.~B.\ 1988, Cambridge  and New York, Cambridge University Press, 1988, 221 p.,   
\bibitem[Vink{\'o} et al.(2004)]{Vinko04} Vink{\'o}, J., Blake, R.~M., S{\'a}rneczky, K., et al.\ 2004, A\&A, 427, 453 
\bibitem[Vink{\'o} et  al.(2012)]{Vinko12} Vink{\'o}, J., Tak{\'a}ts, K., Szalai, T., et al.\ 2012, A\&A, 540, A93  
\bibitem[Woosley \& Bloom(2006)]{WoosleyBloom06} Woosley, S.~E., \& Bloom, J.~S.\ 2006, ARA\&A, 44, 507 

\end{thebibliography}

\appendix

\begin{table*}
\small
\centering
\setlength{\tabcolsep}{10pt}
\caption{SNe IIb, Ib, Ic, IcBL and GRB-SNe: Relavant observables}
\label{table:vitals}
\begin{tabular}{|ccccccc|}
\hline
SN	&	Type	&	$z$	&		$t_{\rm bolo}^{\rm max}$ (day)				&		$m_{\rm B}^{\rm max}$ (mag)				&		$m_{\rm V}^{\rm max}$ (mag)				&		$m_{\rm R}^{\rm max}$ (mag)				\\
\hline																													
2008ax	&	IIb	&	0.0021	&	$	17.65	\pm	1.77	$	&	$	-			$	&	$	-			$	&	$	-			$	\\
2010as	&	IIb	&	0.007354	&	$	13.77	\pm	1.24	$	&	$	-			$	&	$	-			$	&	$	-			$	\\
2011dh	&	IIb	&	0.00155	&	$	17.94	\pm	1.97	$	&	$	-			$	&	$	-			$	&	$	-			$	\\
2011ei	&	IIb	&	0.009317	&	$	18.27	\pm	1.28	$	&	$	-			$	&	$	-			$	&	$	-			$	\\
2011hs	&	IIb	&	0.005701	&	$	15.98	\pm	2.24	$	&	$	-			$	&	$	-			$	&	$	-			$	\\
1999dn	&	Ib	&	0.00938	&	$	13.12	\pm	0.79	$	&	$	-			$	&	$	-			$	&	$	-			$	\\
2005bf$^{**}$	&	Ib	&	0.018913	&	$	39.18	\pm	1.96	$	&	$	-			$	&	$	-			$	&	$	-			$	\\
2008D	&	Ib	&	0.0070	&	$	18.22	\pm	0.55	$	&	$	-			$	&	$	-			$	&	$	-			$	\\
2009jf	&	Ib	&	0.007942	&	$	20.53	\pm	1.64	$	&	$	-			$	&	$	-			$	&	$	-			$	\\
1994I	&	Ic	&	0.00155	&	$	11.20	\pm	0.45	$	&	$	-			$	&	$	-			$	&	$	-			$	\\
2002ap	&	Ic	&	0.002187	&	$	9.89	\pm	0.40	$	&	$	-			$	&	$	-			$	&	$	-			$	\\
2004aw	&	Ic	&	0.0175	&	$	10.13	\pm	0.51	$	&	$	-			$	&	$	-			$	&	$	-			$	\\
2005ek	&	Ic	&	0.016618	&		5.10$^*$				&	$	-			$	&	$	-			$	&	$	-			$	\\
2007gr	&	Ic	&	0.001729	&	$	13.20	\pm	0.53	$	&	$	-			$	&	$	-			$	&	$	-			$	\\
2011bm	&	Ic	&	0.0221	&	$	19.49	\pm	1.17	$	&	$	-			$	&	$	-			$	&	$	-			$	\\
2009bb	&	IcBL	&	0.009987	&	$	12.89	\pm	0.90	$	&	$	-			$	&	$	-			$	&	$	-			$	\\
2012ap	&	IcBL	&	0.012241	&	$	13.15	\pm	1.32	$	&	$	-			$	&	$	-			$	&	$	-			$	\\
1998bw	&	GRB	&	0.00867	&	$	16.12	\pm	0.64	$	&	$	14.01	\pm	0.03	$	&	$	13.50	\pm	0.02	$	&	$	13.49	\pm	0.03	$	\\
2003dh	&	GRB	&	0.1685	&	$	10.73	\pm	1.93	$	&	$	20.36	\pm	0.61	$	&	$	20.02	\pm	0.52	$	&	$	-			$	\\
2003lw	&	GRB	&	0.10536	&	$	19.60	\pm	3.92	$	&	$	-			$	&	$	18.56	\pm	0.24	$	&	$	18.59	\pm	0.24	$	\\
2006aj	&	GRB	&	0.03342	&	$	10.43	\pm	0.52	$	&	$	17.57	\pm	0.06	$	&	$	17.22	\pm	0.05	$	&	$	17.07	\pm	0.07	$	\\
2009nz	&	GRB	&	0.49	&	$	18.55	\pm	3.71	$	&	$	22.59	\pm	0.33	$	&	$	22.39	\pm	0.30	$	&	$	-			$	\\
2010bh	&	GRB	&	0.0592	&	$	9.80	\pm	0.59	$	&	$	-			$	&	$	18.64	\pm	0.08	$	&	$	18.64	\pm	0.08	$	\\
2012bz	&	GRB	&	0.283	&	$	13.38	\pm	1.07	$	&	$	21.70	\pm	0.25	$	&	$	21.26	\pm	0.21	$	&	$	-			$	\\
2013dx	&	GRB	&	0.145	&	$	12.52	\pm	0.63	$	&	$	20.64	\pm	0.24	$	&	$	20.01	\pm	0.24	$	&	$	19.86	\pm	0.24	$	\\
2016jca	&	GRB	&	0.1475	&	$	12.38	\pm	0.62	$	&	$	20.37	\pm	0.20	$	&	$	20.18	\pm	0.20	$	&	$	20.03	\pm	0.20	$	\\
\hline
\end{tabular}
\begin{flushleft}
NB: All times in rest-frame and relative to the explosion epoch (which can be found in table 1 of Paper I).\\
NB: All reference for the SNe listed here can be found in Paper I, expect for GRB-SN~2016jca, which is from \citet{Cano17}.\\
$^{*}$ Properties of the second peak.\\
$^{\dagger}$ Peak unconstrained. Used the time of the first epoch as assumed time of peak.\\
\end{flushleft}
\end{table*}

\bsp

\label{lastpage}

\end{document}